\documentclass[oneside,english,review]{amsart}
\usepackage[latin9]{inputenc}
\usepackage{float}
\usepackage{amsthm}
\usepackage{amstext}
\usepackage{graphicx}
\usepackage{esint}

\makeatletter

\providecommand{\tabularnewline}{\\}

\numberwithin{equation}{section}
\numberwithin{figure}{section}

{\tiny{}}\global\long\def\bra#1{\Bra{#1}}
{\tiny{}}\global\long\def\ket#1{\Ket{#1}}
{\tiny{}}
{\tiny{}}\global\long\def\braket#1{\Braket{#1}}
{\tiny \par}

\usepackage{lineno}
\usepackage{amstext}
\usepackage{braket}
\usepackage{algorithm}
\usepackage{algorithmicx}
\usepackage{soul}
\usepackage{verbatim}
\usepackage{newunicodechar}

\usepackage[english]{babel}

\modulolinenumbers[5]

\date{}

\makeatother

\usepackage{babel}
\begin{document}
\date{}

\title{Numerical implementation of the multiscale and averaging methods
for quasi periodic systems }

\author{Tal Kachman}

\address{Department of Physics, Physics of Living Systems Group, Massachusetts
Institute of Technology, Floor 6, 400 Tech Square, Cambridge, MA 02139,\\
Physics Department, Technion- Israel Institute of Technology, Haifa
3200, Israel}

\email{kachman@mit.edu}

\author{Shmuel Fishman}

\address{Physics Department, Technion- Israel Institute of Technology, Haifa
3200, Israel}

\author{Avy Soffer}

\address{Mathematics Department, Rutgers University, New-Brunswick, NJ 08903,
USA}
\begin{abstract}
We consider the problem of numerically solving the Schrödinger equation
with a potential that is quasi periodic in space and time. We introduce
a numerical scheme based on a newly developed multi-time scale and
averaging technique. We  demonstrate that with this novel method we
can solve efficiently and with rigorous control of the error such
an equation for long times. A comparison with the standard split-step
method shows substantial improvement in computation times, besides
the controlled errors. We apply this method for a free particle driven
by quasi-periodic potential with many frequencies. The new method
makes it possible to evolve the Schr\"{o}dinger equation for times
much longer than was possible so far and to conclude that there are
regimes where the energy growth stops in-spite of the driving.
\end{abstract}

\maketitle



\section{Introduction}

A method for the study of the dynamics for the Schrödinger equation
with time dependent potentials \cite{fishman2014multiscale} is implemented
numerically. The potential is quasiperiodic in both space and time.
The power of this new method is demonstrated. Exploration of the dynamics
for such potentials is motivated by experiments in optics \cite{levi2012hyper}
where hypertransport, namely transport faster then ballistic was found
experimentally and numerically in some regimes. In theoretical work
that followed \cite{levi2012hyper,krivolapov2012super,krivolapov2012universality,krivolapov2012transport}
a classical theory was developed for the potentials relevant for these
optics experiments. In particular, it was found in the framework of
classical mechanics, that for smooth potentials the spreading in momentum
as function of time stops. The calculations of the present paper are
for such potentials. For short times, relevant for the existing experiments
it was found that wave or quantum and classical dynamics agree in
general features.\\
For long times it turned out impossible to compute numerically the
quantum dynamics using the standard methods \cite{maday1990operator,kosloff1988time},
while with the method introduced in \cite{fishman2014multiscale}
and implemented here, calculations for such long times are feasible
as will be demonstrated in this paper. Such calculations and comparison
with the classical results, is of fundamental importance for the issue
of quantum classical correspondence. The main objective of the present
paper is to demonstrate the power of the method introduced in \cite{fishman2014multiscale}
for a physically relevant example.

Potentials which are quasiperodic both in space and time, can manifest
a high degree of complexity and are subject of many studies over the
last decade, mostly in the framework of classical physics \cite{PhysRev.36.823,PhysRev.141.186,gardiner1985stochastic,xx1,PhysRevLett.96.030601,PhysRevLett.69.1831,PhysRevLett.67.2115}.
With different experimental realizations of such potentials there
is also a need for a numerical approach to investigate them and their
asymptotic behavior. The standard way to numerically solve problems
with time dependent potentials is based on either spectral methods
\cite{hesthaven2007spectral,leforestier1991comparison} or explicit/implicit
finite difference schemes \cite{kosloff1988time,ascher1995implicit}.
In this paper we implement numerically a recently developed rigorously
controlled multi-time scale averaging technique \cite{fishman2014multiscale}.
The above mentioned method has two distinct advantages. The first
is that at each step of the averaging hierarchy there is a well defined
and completely known bound on the numerical error. The second advantage
and one which has far greater impact is the reduction in computational
time accompanied with each of the hierarchical averaging steps. This
reduction enables us to go to long time scales, impossible by the
methods we are aware of. In this paper we describe and present an
implementation for a specific problem.

The outline of the paper is as follows. In Section 2, we present the
model for which the method of multiple scale and averaging (MSA) is
implemented. This model is of physical interest and importance. The
MSA method that was developed in Ref. \cite{fishman2014multiscale}
and details of its implementation are presented in Secs. 3 and 4.
The MSA method involves a hierarchy of computations where the first
level is described is Sec. 3 and the following levels are presented
in Section 4. The needed level is determined by the required precision
and the time over which the system is evolved. In Section 5 we demonstrate
that the MSA method is superior compared to the standard split step
method, since for the same precision it is much faster. Moreover for
the split step method there is only an empirical estimate on the error
while for the MSA there is a rigorous bound. Using it in Section 6,
we show that with the help of this method we could solve the Schr\"{o}dinger
equation for the model problem presented in Section 2, for an extremely
long time; we conclude that for this model the energy does not grow
to infinity in-spite of the time dependent driving potential. The
spreading in the quantum case is wider than in the corresponding classical
system. This results from the fact that initially we observe a lot
of spreading in the quantum case, while the classical case does not
show spreading.

\section{\label{sec:The-model-that}The model that will be studied }

In this section we introduce the model for which our MSA method developed
in \cite{fishman2014multiscale} is implemented. In the first subsection
its relation to physical systems is explained, while in the second
one it is reduced to a form for which the MSA method can be applied(see
Eqs. \ref{eq:random potential in moving frame}-\ref{schred_pertr}).

\subsection{The physical model}

The random potentials which are prepared in optics \cite{schwartz2007transport,levi2012hyper}
and atom optics \cite{piraud2011localization,billy2008direct,sanchez2007anderson}
experiments, are described by a sum of random Fourier components.
In experiments, potentials which are composed out of a large number
of random independent Fourier components $N$, are created . \\
More specifically here, the Schr\"{o}dinger equation for a potential
\begin{equation}
V\left(x',\tau\right)=\frac{1}{\sqrt{N}}\sum_{n=1}^{N}A_{n}\exp\left(i(k_{n}x'-\omega'_{n}\tau)\right)+\mathcal{C}.\mathcal{C}\label{pot}
\end{equation}
is used. The $A_{m}$ are independent, identically distributed complex
random variables, where $\mathcal{C}.\mathcal{C}$ stands for complex
conjugate. The expectation values of these variables satisfy 
\begin{equation}
<A_{m}>=<A_{m}A_{n}>=0\ \ \ <A_{m}{A_{n}}^{*}>=\sigma^{2}\delta_{nm}
\end{equation}
We will study the specific model where $A_{n}=A\ e^{i\phi_{m}}$ with
$\phi_{m}$ uniformly distributed in the interval $[-\pi,\pi]$ and
$A>0$. The distribution of $\omega_{n}$ and $k_{n}$ is specified
in Section 5. \\
The equation of motion is the time dependent Schrödinger equation

\begin{equation}
i\partial_{\tau}\psi\left(x',\tau\right)=H\psi\left(x',\tau\right).\label{SE}
\end{equation}
Where 
\begin{equation}
H=\frac{1}{2}p'^{2}+V(x',\tau),
\end{equation}
\[
p'=-i\nabla_{x'}.
\]
In the following section we will introduce the reduction of the problem
to a form where the MSA technique is applicable.

\subsection{Reduction of the problem }

In the model with potential given by (\ref{pot}), the particle is
expected to be accelerated most effectively in the regime of velocities
where the Chirikov resonances 
\begin{equation}
v_{n}^{(r)}=\frac{\omega'_{n}}{k_{n}}\label{eq:5_w}
\end{equation}
are formed \cite{Chirikov1979263,Zaslavski}. \\
In this paper we choose $v_{n}^{(r)}$ and $k_{n}$ uniformly distributed
in the intervals $\left[v^{\left(\text{min}\right)},v^{\left(\text{max}\right)}\right]$
and $\left[k^{\left(\text{min}\right)},k^{\left(\text{max}\right)}\right]$
respectively. The crucial point is that the Chirikov resonant velocities
are bounded in a phase space strip. This is typically the case for
smooth potentials, and will be assumed in this paper. \\
Here we would like to study what is the acceleration of a particle
prepared with momentum or velocity $v$ (we assume unit mass), so
that all $v_{n}^{(r)}$ are far from $v$. For the classical corresponding
system we found that the acceleration is negligible \cite{krivolapov2012super,krivolapov2012transport,krivolapov2012universality}.
Here we study the corresponding quantum mechanical system. For this
purpose it is convenient to work in a frame of reference where the
initial velocity of the particle vanishes. For this we perform the
Galilean transformation 
\begin{equation}
x=x'-v\tau,
\end{equation}
where $v$ is the velocity of the moving frame (in later stages we
will relate this quantity to the required small parameter) on the
potential (\ref{pot}) 
\begin{eqnarray}
V\left(x+v\tau,\tau\right)=\frac{A}{\sqrt{N}}\sum_{n=1}^{N}\cos\left(k_{n}\left(x+v\tau\right)-\omega'_{n}\tau+\phi_{n}\right)\\
=\frac{A}{\sqrt{N}}\sum_{n=1}^{N}\cos\left(k_{n}x+\left(k_{n}v-\omega'_{n}\right)\tau+\phi_{n}\right).\nonumber 
\end{eqnarray}
Now, re-scaling time as 
\begin{equation}
t=\tau v,\label{tau_scale}
\end{equation}

the potential takes the form 
\begin{equation}
V\left(x,t\right)=\frac{A}{\sqrt{N}}\sum_{n=1}^{N}\cos\left(k_{n}x-\frac{\left(\omega'_{n}-k_{n}v\right)}{v}t+\phi_{n}\right)
\end{equation}
and defining 
\begin{equation}
\omega{}_{n}=\frac{\left(\omega'_{n}-k_{n}v\right)}{v}
\end{equation}
The potential in the moving frame takes the form 
\begin{equation}
V\left(x,t\right)=\frac{A}{\sqrt{N}}\sum_{n=1}^{N}\cos\left(k_{n}x-\omega{}_{n}t+\phi_{n}\right)\label{eq:random potential in moving frame}
\end{equation}
The time dependent Schrödinger equation is 
\begin{equation}
i\partial_{t}\psi\left(x,t\right)=\frac{1}{v}H(p,x,t)\psi\left(x,t\right)
\end{equation}

Introducing the small parameter $\beta=\frac{1}{v}$: 
\begin{equation}
i\partial_{t}\psi\left(x,t\right)=\beta H(p,x,t)\psi\left(x,t\right)\label{schred_pertr}
\end{equation}
From this point on we will use the small parameter $\beta$ to perform
the averaging steps introduced in the next section. In what follows
we will also relate the small parameter $\beta$ to the time scale
on which we average. The Hamiltonian will be approximated by a finite
matrix (in space and momentum) and it will be verified that the spreading
never reaches the boundaries set by this basis. \

\section{The averaging scheme}

The multiscale averaging method is based on replacing the original
Hamiltonian by a hierarchical set of averaged Hamiltonians. In each
step we perform a ``peel-off'' transformation and average a part
of the Hamiltonian, for a chosen time interval of length $T_{0}=\frac{1}{\sqrt{\beta}}=\sqrt{v}$.
\ This choice is not unique, but as shown in \cite{fishman2014multiscale}
leads to effective error bounds. We use the fact that Eq (\ref{schred_pertr})
is of the form of (2.1) in \cite{fishman2014multiscale}.\\

In this section the implementation of the MSA method of \cite{fishman2014multiscale}
for equation (\ref{schred_pertr}) will be presented. In Appendix
A we summarize the main results of Ref. \cite{fishman2014multiscale}

\subsection{Zero order}

The zero order average on the jth time interval has the form ($V\left(t\right)\equiv V\left(x,t\right)$)
\begin{equation}
\bar{V}_{0}^{\left(j\right)}=\frac{1}{T_{0}}\int_{jT_{0}}^{(j+1)T_{0}}V\left(t\right)dt
\end{equation}
In the case of the potential (\ref{pot}), the zero order averaging
can be performed analytically 
\begin{eqnarray}
\bar{V}_{0}^{\left(j\right)} & = & \frac{1}{T_{0}}\int_{jT_{0}}^{(j+1)T_{0}}V\left(t\right)dt=\frac{A}{\sqrt{N}T_{0}}\int_{jT_{0}}^{(j+1)T_{0}}\sum_{n=1}^{N}\cos\left(k_{n}x-\omega{}_{n}t+\phi_{n}\right)=\nonumber \\
 & = & \frac{A}{\sqrt{N}T_{0}}\sum_{n=1}^{N}\frac{1}{\omega}\left[\sin\left(k_{n}x-\omega{}_{n}\left(j+1\right)T_{0}+\phi_{n}\right)\right.\nonumber \\
 &  & \left.-\sin\left(k_{n}x-\omega{}_{n}jT_{0}+\phi_{n}\right)\right]=\nonumber \\
 & = & 2\frac{A}{\sqrt{N}T_{0}}\sum_{n=1}^{N}\frac{1}{\omega_{n}}\sin\left(\frac{1}{2}\omega{}_{n}T_{0}\right)\cos\left(k_{n}x-\omega{}_{n}\left(j+\frac{1}{2}\right)T_{0}+\phi_{n}\right)\label{eq:zero order average}
\end{eqnarray}
This defines the Hamiltonian on one time interval; accordingly we
can write the Hamiltonian of one interval as 
\begin{equation}
\bar{H}_{0}^{(j)}\left(x\right)=-\frac{1}{2}\frac{\partial^{2}}{\partial x^{2}}+\bar{V}_{0}^{\left(j\right)}\left(x\right)\label{eq:H_0}
\end{equation}
The global Hamiltonian corresponding to (2.1) of \cite{fishman2014multiscale},
which gives the zeroth order approximation to $H\left(t\right)$,
is generated by: 
\begin{equation}
\bar{H}_{0}^{g}={\bar{H}_{0}}^{(j)}(t)\ \ \text{for}\ \ jT_{0}\leq t<(j+1)T_{0}.
\end{equation}
Using this notation, a general time evolution can be written as the
product of the interval propagators since $\bar{H}_{0}^{g}$ is piecewise
constant in time,

\begin{equation}
U_{0}\left(t\right)=e^{-i\beta\bar{H}_{0}^{(j_{max})}\left(t-j_{max}T_{0}\right)}\dots e^{-i\beta\bar{H}_{0}^{(1)}T_{0}}e^{-i\beta\bar{H}_{0}^{(0)}T_{0}}\label{eq:3.5}
\end{equation}
 where $j_{max}-1$ is the integer part of $t/T_{0}$ The evolution
in this order is 
\begin{equation}
\psi_{0}\left(x,t\right)=U_{0}\left(t\right)\psi\left(x,0\right).\label{zero_order_prog}
\end{equation}
The propagator satisfies 
\begin{equation}
i\frac{\partial}{\partial t}U_{0}(t)=\beta\bar{H}_{0}^{(g)}U_{0}(t)\label{eq:20}
\end{equation}
corresponding to (2.10) of \cite{fishman2014multiscale}. This propagator
is numerically implemented and used to solve the time dependent equation
of motion.  The error in this order is bounded by  $\beta^{\frac{1}{2}}$
up to times of order $T_{0}$, the reasoning behind this is shown
in \cite{fishman2014multiscale}, Eq. (2.49) there. The error in diagonalization
of $\bar{H}_{0}^{\left(j\right)}$ is negligible (see discussion in
Sec. 5)

\subsection{First order\label{first_order_scheme}}

The first order averaging is based upon a ``peel-off'' transformation
of the zero order. In such a transformation the next order Hamiltonian
is constructed from the zero order in the following way: Let $H_{1}\left(t\right)$
be defined as 
\begin{equation}
H_{1}\left(t\right)=U_{0}^{-1}\left(t\right)\left[H\left(t\right)-\bar{H_{0}}^{g}\left(t\right)\right]U_{0}\left(t\right)
\end{equation}
Hence $H_{1}\left(t\right)$ is the Heisenberg dynamics of the full
problem with $U_{0}\left(t\right)$ dynamics peeled off. An important
point to note is that the Laplacian term drops in $H_{1}!$ Hence
$H_{1}\left(t\right)$ is a bounded operator, for which the results
of {[}1{]} directly apply. Its average in the j-th interval is 
\begin{equation}
\bar{H}_{1}^{\left(j\right)}=\frac{1}{T_{0}}\int_{jT_{0}}^{\left(j+1\right)}H_{1}\left(t\right)dt
\end{equation}
 The $\bar{H}_{1}\left(t\right)$ dynamics is given by the propagator
\begin{equation}
U_{1}\left(t\right)=e^{-i\beta\bar{H}_{1}^{(j_{max})}\left(t-j_{max}T_{0}\right)}\dots e^{-i\beta\bar{H}_{1}^{(0)}T_{0}}\label{first_order_prop}
\end{equation}
We turn now to calculate $\bar{H}_{1}^{\left(j\right)}$, by (2.49)
of \cite{fishman2014multiscale} it is of order $\sqrt{\beta}$. Before
diagonalizing this operator  in order to calculate the time evolution,
there are several steps needed to be taken. First we write explicitly
$\bar{H_{1}}^{\left(j\right)}$ using integration by parts 
\begin{equation}
\bar{H_{1}}^{\left(j\right)}=\frac{1}{T_{0}}\int_{jT_{0}}^{(j+1)T_{0}}U_{0}^{-1}(t)\left[H\left(t\right)-\bar{H}_{0}^{g}\left(t\right)\right]U_{0}(t)dt=\bar{H_{1}}^{\left(j,\text{I}\right)}+\bar{H_{1}}^{\left(j,\text{II}\right)}+\bar{H_{1}}^{\left(j,\text{III}\right)},\label{eq:24}
\end{equation}
where 
\begin{equation}
\bar{H_{1}}^{\left(j,\text{I}\right)}=\frac{1}{T_{0}}\left[U_{0}^{-1}(t)\left(\int_{0}^{t}\left[H\left(t'\right)-\bar{H}_{0}^{g}\left(t'\right)\right]dt^{\prime}\right)U_{0}(t)\right]\begin{array}{c}
(j+1)T_{0}\\
t=jT_{0}
\end{array}\label{eq:2.34}
\end{equation}
\begin{equation}
\bar{H_{1}}^{\left(j,\text{II}\right)}=-\frac{1}{T_{0}}\int_{nT_{0}}^{(n+1)T_{0}}dt^{\prime}\left(\frac{\partial}{\partial t^{\prime}}U_{0}^{-1}(t^{\prime})\right)\left[\int_{0}^{t^{\prime}}\left[H\left(s\right)-\bar{H}_{0}^{g}\left(s\right)\right]ds\right]U_{0}(t^{\prime}),\label{eq:26}
\end{equation}

and 
\begin{equation}
\bar{H_{1}}^{\left(j,\text{III}\right)}=-\frac{1}{T_{0}}\int_{jT_{0}}^{(j+1)T_{0}}U_{0}^{-1}(t^{\prime})\left[\int_{0}^{t^{\prime}}\left[H\left(s\right)-\bar{H}_{0}^{g}\left(s\right)\right]ds\right]\left(\frac{\partial}{\partial t}U_{0}\right)(t^{\prime})dt^{\prime}.
\end{equation}
We note that for any integer $t/T_{0}$, 
\begin{equation}
\int_{jT_{0}}^{\left(j+1\right)T_{0}}\left[H\left(t'\right)-\bar{H}_{0}^{g}\left(t'\right)\right]dt'=0\label{zero_hamil}
\end{equation}
by construction. Therefore for such $t$ $\bar{H_{1}}^{\left(j,\text{I}\right)}\left(t\right)=0$
and $\bar{H_{1}}^{\left(j,\text{II}\right)}\left(t\right)$ can be
simplified. Moreover the expression in $\bar{H_{1}}^{\left(j,\text{III}\right)}$
is just the hermitian conjugate of $\bar{H_{1}}^{\left(j,\text{II}\right)}$
so we only need to analyze the second term: 
\begin{eqnarray}
\bar{H_{1}}^{\left(j,\text{II}\right)} & = & \frac{1}{T_{0}}\int_{jT_{0}}^{(j+1)T_{0}}dt^{\prime}\left(\frac{\partial}{\partial t^{\prime}}U_{0}^{-1}(t^{\prime})\right)\left[\int_{jT_{0}}^{t^{\prime}}\left[H_{j}\left(s\right)-\bar{H_{0}}^{\left(g\right)}\left(s\right)\right]ds\right]U_{0}(t^{\prime})\label{eq:29}\\
 &  & =\frac{1}{T_{0}}\int_{jT_{0}}^{(j+1)T_{0}}dt^{\prime}\left(\frac{\partial}{\partial t^{\prime}}U_{0}^{-1}(t^{\prime})\right)\left[\int_{jT_{0}}^{t^{\prime}}H_{0}\left(s\right)ds-\left(t'-jT_{0}\right)\bar{H_{0}}^{\left(j\right)}\right]U_{0}(t^{\prime})\nonumber 
\end{eqnarray}
To evaluate this operator we can use a recursive construction of states
based on the zeroth order eigenstates of the averaged Hamiltonian.
It is important to remember that this is actually a matrix. For convenience
of notation we will define 
\begin{equation}
U_{0}\left(t\right)=e^{-i\beta\bar{H}_{0}^{(j)}\left(t-jT_{0}\right)}......e^{-i\beta\bar{H}_{0}^{\left(1\right)}T_{0}}e^{-i\beta\bar{H}_{0}^{\left(0\right)}T_{0}},
\end{equation}
For $t$ that is an integer multiple of $T_{0}$. 
\begin{equation}
U_{0}\left(t\right)=\prod_{j'=0}^{j}W_{j-j'}\label{eq:31}
\end{equation}
 where 
\begin{equation}
W_{j}(t)=e^{-i\beta\bar{H}_{0}^{\left(j\right)}T_{0}}.\label{WJ}
\end{equation}
First we assume $t/T_{0}$ is integer and then generalize for any
arbitrary $t$. The eigenvalues and eigenfunctions of $\bar{H}_{0}^{\left(j\right)}$
are 
\begin{equation}
\bar{H}_{0}^{\left(j\right)}\ket{\varphi_{j}^{k}}=E_{j}^{k}\ket{\varphi_{j}^{k}}\label{diag_zero}
\end{equation}
Here we use a base of size $M$ that is assumed to be finite (in this
work we take $M=64$). Since $\beta$ is small the eigenfunctions
are approximately eigenfunctions of a free particle, namely 
\begin{equation}
<x|\varphi_{j}^{k}>\approx e^{i\tilde{k}x}.\label{eq:basis_wave}
\end{equation}
 We will have $M$ eigenvalues and eigenvectors, the largest value
of $\left|\tilde{k}\right|$ is 50. Between each pair of propagators
$W_{j},W_{i-1}$ we can insert the identity resolution in the corresponding
basis (\ref{diag_zero}) 
\begin{equation}
\hat{I}_{j}=\sum_{k=1}^{M}\ket{\varphi_{j}^{k}}\bra{\varphi_{j}^{k}}\label{eq:res_idn}
\end{equation}
resulting in 
\begin{equation}
U_{0}(t)=W_{j}\cdots W_{1}W_{0}=W_{j}\hat{I_{j}}\cdots W_{1}I_{1}W_{0}I_{0}
\end{equation}
As an example let us take just the first two terms 
\begin{equation}
W_{1}I_{1}W_{0}I_{0}=\sum_{k_{1}=1}^{M}W_{1}\ket{\varphi_{1}^{k_{1}}}\bra{\varphi_{1}^{k_{1}}}\sum_{k_{0}=1}^{M}W_{0}\ket{\varphi_{0}^{k_{0}}}\bra{\varphi_{0}^{k_{0}}}
\end{equation}
the brakets expressions $\bra{\varphi_{1}^{k_{1}}}\sum_{k_{0}=1}^{M}W_{0}\ket{\varphi_{0}^{k_{0}}}$
are just scalars, and since $\varphi_{j}^{k_{j}}$ are eigenfunctions
of $H_{0}^{j}$ \eqref{diag_zero} 
\begin{eqnarray}
W_{1}I_{1}W_{0}I_{0} & = & \sum_{k_{1}=1}^{M}W_{1}\ket{\varphi_{1}^{k_{1}}}\bra{\varphi_{1}^{k_{1}}}W_{0}\sum_{k_{0}=1}^{M}\ket{\varphi_{0}^{k_{0}}}\bra{\varphi_{0}^{k_{0}}}\\
 & = & \sum_{k_{1}=1}^{M}e^{-i\beta T_{0}E_{1}^{k_{1}}}\ket{\varphi_{1}^{k_{1}}}\bra{\varphi_{1}^{k_{1}}}\sum_{k_{0}=1}^{M}e^{-i\beta T_{0}E_{0}^{k_{0}}}\ket{\varphi_{0}^{k_{0}}}\bra{\varphi_{0}^{k_{0}}}\nonumber 
\end{eqnarray}
using the notation 
\begin{equation}
\alpha_{k_{1}k_{0}}=\braket{\varphi_{1}^{k_{1}}|\varphi_{0}^{k{}_{0}}}\label{alphas}
\end{equation}
this becomes 
\begin{equation}
W_{1}\hat{I}_{1}W_{0}\hat{I}_{0}=\sum_{k_{0}=1}^{M}\sum_{k_{1}=1}^{M}\alpha_{k_{1}k_{0}}e^{-i\beta T_{0}E_{1}^{k_{1}}}e^{-i\beta T_{0}E_{0}^{k_{0}}}\ket{\varphi_{1}^{k_{1}}}\bra{\varphi_{0}^{k_{0}}}
\end{equation}
In the same way for the complete sequence of propagators \eqref{eq:31}
\begin{eqnarray}
U_{0}\left(t\right)= &  & \sum_{k_{0}=1}^{M}\dots\sum_{k_{j}=1}^{M}\alpha_{k_{j}k_{j-1}}\alpha_{k_{j-1}k_{j-2}}\dots\nonumber \\
 &  & \dots\alpha_{k_{1}k_{0}}e^{-i\beta\left(t-jT_{0}\right)E_{j}^{k_{j}}}\dots e^{-i\beta T_{0}E_{1}^{k_{1}}}e^{-i\beta T_{0}E_{0}^{k_{0}}}\ket{\varphi_{j}^{k_{j}}}\bra{\varphi_{0}^{k_{0}}}.\nonumber \\
\label{zero_order_prop}
\end{eqnarray}
and it satisfies \eqref{eq:20}. \\
Here $jT_{0}<t<\left(j+1\right)T_{0}$.

The inverse takes the form 
\begin{eqnarray}
U_{0}^{-1}(t) & = & e^{i\beta\bar{H}_{0}^{\left(0\right)}T_{0}}e^{i\beta\bar{H}_{0}^{\left(1\right)}T_{0}}......e^{i\beta\bar{H}_{0}\left(t-jT_{0}\right)}\\
 & = & \sum_{k_{j}=1}^{M}\dots\sum_{k_{j-1}=1}^{M}\alpha_{k_{0}k_{1}}^{*}\alpha_{k_{1}k_{2}}^{*}e^{i\beta\left(t-jT_{0}\right)E_{0}}e^{i\beta\left(j-1\right)E_{0}}\ket{\varphi_{0}^{k_{0}}}\bra{\varphi_{j}^{k_{j}}}\nonumber 
\end{eqnarray}
By \eqref{eq:20} or direct differentiation of \eqref{zero_order_prop}
\begin{eqnarray}
\frac{\partial}{\partial t}U_{0}(t) & = & -\sum_{k_{0}^{'}=1}^{M}\dots\sum_{k'_{j}=1}^{M}\sum_{k_{0}=1}^{M}\dots\sum_{k_{j}=1}^{M}i\beta E_{j}^{k_{j}}\alpha_{k_{j}k_{j-1}}\alpha_{k_{j-1}k_{j-2}}\dots\alpha_{k_{1}k_{0}}\label{eq:43}\\
 &  & e^{-i\beta\left(t-jT_{0}\right)E_{j}^{k_{j}}}\dots e^{-i\beta T_{0}E_{0}^{k_{0}}}\ket{\varphi_{j}^{k_{j}}}\bra{\varphi_{0}^{k_{0}}}\nonumber 
\end{eqnarray}
and in a similar way one finds 
\begin{equation}
\frac{\partial}{\partial t}U_{0}^{-1}(t)=\sum_{k_{j}=1}^{M}\dots\sum_{k_{j-1}=1}^{M}\alpha_{k_{0}k_{1}}^{*}\alpha_{k_{1}k_{2}}^{*}\dots\alpha_{k_{j-1}k_{j}}^{*}e^{i\beta T_{0}E_{0}^{k_{0}}}\dots e^{i\beta\left(t-jT_{0}\right)E_{j}^{k_{j}}}i\beta E_{j}^{k_{j}}\ket{\varphi_{0}^{k_{0}}}\bra{\varphi_{j}^{k_{j}}}
\end{equation}
substitution of \eqref{eq:43} and \eqref{zero_order_prop} into
\eqref{eq:29} leads to  
\begin{eqnarray}
\bar{H_{1}}^{\left(j,\text{II}\right)}\left(t\right) & = & -\frac{1}{T_{0}}\sum_{k_{0}^{'}=1}^{M}\dots\sum_{k'_{j}=1}^{M}\sum_{k_{0}=1}^{M}\dots\sum_{k_{j}=1}^{M}\beta E_{j}^{k_{j}}\alpha_{k'_{0}k'_{1}}^{*}\dots\alpha_{k'_{j-1}k'_{j}}^{*}\alpha_{k{}_{j}k{}_{j-1}}\dots\alpha_{k_{1}k_{0}}\nonumber \\
 &  & \int_{jT_{0}}^{(j+1)T_{0}}dt^{\prime}e^{i\beta T_{0}E_{0}^{k'_{0}}}e^{-i\beta T_{0}E_{0}^{k'_{0}}}\dots e^{i\beta\left(t-jT_{0}\right)E_{j}^{k'_{j}}}e^{-i\beta\left(t'-jT_{0}\right)E_{j}^{k'_{j}}}\nonumber \\
 &  & \ket{\varphi_{0}^{k_{0}^{\prime}}}\bra{\varphi_{j}^{k_{j}^{\prime}}}\left[\int_{jT_{0}}^{t^{\prime}}H\left(s\right)ds-\left(t'-jT_{0}\right)\bar{H_{0}}^{\left(j\right)}\right]\ket{\varphi_{j}^{k_{j}}}\bra{\varphi_{0}^{k_{0}}}=\nonumber \\
 & = & -\frac{1}{T_{0}}\sum_{k_{0}^{'}=1}^{M}\dots\sum_{k'_{j}=1}^{M}\sum_{k_{0}=1}^{M}\dots\sum_{k_{j}=1}^{M}\beta E_{j}^{k_{j}}\alpha_{k'_{0}k'_{1}}^{*}\dots\alpha_{k'_{j-1}k'_{j}}^{*}\alpha_{k{}_{j}k{}_{j-1}}\dots\alpha_{k_{1}k_{0}}\nonumber \\
 &  & \int_{jT_{0}}^{(j+1)T_{0}}dt'e^{-i\beta T_{0}E_{0}^{k'_{0}}}\dots e^{i\beta\left(t-jT_{0}\right)E_{j}^{k'_{j}}}e^{-i\beta\left(t-jT_{0}\right)E_{j}^{k'_{j}}}\nonumber \\
 &  & \left\langle \left.\varphi_{j}^{k_{j}}\right|\int_{jT_{0}}^{t^{\prime}}H\left(s\right)ds-\left(t-jT_{0}\right)\bar{H_{0}}^{\left(j\right)}\left|\varphi_{j}^{k'_{j}}\right.\right\rangle \ket{\varphi_{0}^{k'_{0}}}\bra{\varphi_{0}^{k_{0}}}
\end{eqnarray}
Finally the full expression for the first order averaged Hamiltonian
is, where \eqref{eq:24} and \eqref{zero_hamil} were used, 
\begin{equation}
\bar{H_{1}}^{\left(j\right)}\left(t\right)=\bar{H_{1}}^{\left(j,\text{II}\right)}\left(t\right)+\left(\bar{H_{1}}^{\left(j,\text{II}\right)}\left(t\right)\right)^{\dagger}.
\end{equation}
\begin{eqnarray}
\bar{H_{1}}^{\left(j\right)}\left(t\right) & = & 2\Re\left\{ -\frac{1}{T_{0}}\sum_{k_{0}^{'}=1}^{M}\dots\sum_{k'_{j}=1}^{M}\sum_{k_{0}=1}^{M}\dots\sum_{k_{j}=1}^{M}\beta E_{j}^{k_{j}}\alpha_{k'_{0}k'_{1}}^{*}\dots\alpha_{k'_{j-1}k'_{j}}^{*}\right.\nonumber \\
 &  & \alpha_{k{}_{j}k{}_{j-1}}\dots\alpha_{k_{1}k_{0}}\int_{jT_{0}}^{(j+1)T_{0}}dt'e^{-i\beta T_{0}E_{0}^{k'_{0}}}\dots e^{i\beta\left(t-jT_{0}\right)E_{j}^{k'_{j}}}e^{-i\beta\left(t-jT_{0}\right)E_{j}^{k'_{j}}}\label{eq:H_1_j}\\
 &  & \left.\left\langle \left.\varphi_{j}^{k_{j}}\right|\int_{jT_{0}}^{t^{\prime}}dsH\left(s\right)-\left(t-jT_{0}\right)\bar{H_{0}}^{\left(j\right)}\left|\varphi_{j}^{k'_{j}}\right.\right\rangle \right\} \ket{\varphi_{0}^{k'_{0}}}\bra{\varphi_{0}^{k{}_{0}}}\nonumber 
\end{eqnarray}

The integral \eqref{eq:H_1_j} is performed numerically as follows.
The domain of integration is derived into squares of size $\Delta t\times\Delta t$.
The integral is approximated by a sum of the values of the integrand
of the middle points of the squares, multiplied by $\Delta t^{2}$.
The error in each term is of the order of $\Delta t^{2}$. This can
be improved substantially. Here it is not required since we can obtain
the required precision in this simple way.

At first sight this expression \eqref{eq:H_1_j} might seem very complicated
but can be understood quite easily; in fact what we have here is a
matrix constructed from the sum of matrix products; the terms of the
matrix involve only the products of eigenvalues and eigenvectors of
$\bar{H}_{0}^{\left(j\right)}$. The benefit of calculating the propagator
in this manner is that one only needs to diagonalize the Hamiltonian
in each interval only once. As a result of \eqref{eq:zero order average}
the averaged potential is of order $\sqrt{\beta}$ and $\alpha_{k_{j},k_{i}}$
are elements of matrices that are almost diagonal as will be verified
aposteriori in Sec. 5. Consequently the error in the calculation of
\eqref{eq:H_1_j} is of the order 
\begin{equation}
\delta I_{t}=\Delta t^{2}\label{eq:del_I_t}
\end{equation}
 By the general theory see (2.15) of \cite{fishman2014multiscale}
the first order propagator takes the form \eqref{eq:3.5} with $\bar{H}_{0}^{\left(j\right)}$
replaced by $H_{1}^{\left(j\right)}$ namely with 
\[
U_{1}=e^{-i\beta\bar{H}_{1}^{(j)}\left(t-jT_{0}\right)}......e^{-i\beta\bar{H}_{1}^{\left(1\right)}T_{0}}e^{-i\beta\bar{H}_{1}^{\left(0\right)}T_{0}}
\]

\subsection{Normal form transformation}

To improve the accuracy we implement the normal form transformation;
we will use the form given in Eq. (3.10) of \cite{fishman2014multiscale}.
In our notation this becomes 
\begin{equation}
\tilde{U}_{1}\left(t\right)=1+i\beta U_{1}^{-1}\left(t\right)\left(\int_{0}^{t}dt'\left[H_{1}\left(t'\right)-\bar{H}_{1}^{\left(g\right)}\left(t'\right)\right]\right)U_{1}\left(t\right).
\end{equation}
The manner in which we will simplify the above expression will be
similar to the method used to calculate $U_{1}$ and $H_{1}$ given
in Eq (\ref{first_order_prop}) and (20), splitting into intervals
of length $T_{0}$ and using (\ref{zero_hamil}), replacing $H_{0}$
by $H_{1}$: 
\begin{eqnarray}
\tilde{U}_{1}\left(t\right) & = & 1+i\beta U_{1}^{-1}\left(\sum_{j'=0}^{j}\int_{j'T_{0}}^{\left(j'+1\right)T_{0}}dt'\left[H_{1}\left(t'\right)-\bar{H}_{1}^{\left(j'\right)}\right]\right.\\
 &  & \left.+\int_{jT_{0}}^{t}\left[H_{1}\left(t'\right)+\bar{H}_{1}^{\left(j\right)}\left(t'\right)\right]dt'\right)U_{1}\left(t\right).\nonumber 
\end{eqnarray}
by definition of $\bar{H}_{1}^{\left(j\right)}$ 
\[
\int_{j'T_{0}}^{\left(j'+1\right)T_{0}}dt'\left[H_{1}\left(t'\right)-\bar{H}_{1}^{\left(j'\right)}\right]=0
\]
 
\begin{eqnarray}
\tilde{U}_{1}\left(t\right) & = & 1+i\beta U_{1}^{-1}\int_{jT_{0}}^{t}dt'\left[H_{1}\left(t'\right)-\bar{H}_{1}^{\left(j\right)}\left(t'\right)\right]U_{1}\left(t\right)=\\
 &  & 1+i\beta U_{1}^{-1}\left[\left(\int_{jT_{0}}^{t}H_{1}\left(t'\right)dt'\right)-\left(t-jT_{0}\right)\bar{H}_{1}^{\left(j\right)}\right]U_{1}\left(t\right)\nonumber 
\end{eqnarray}
and explicitly 
\begin{equation}
\tilde{U}\left(t\right)=1+i\beta e^{i\beta\bar{H}_{1}^{\left(0\right)}T_{0}}\dots e^{i\beta\bar{H}_{1}^{\left(j\right)}\left(t-jT_{0}\right)}\left[\left(\int_{jT_{0}}^{t}H_{1}\left(t'\right)dt'\right)-\left(t-jT_{0}\right)\bar{H}_{1}^{\left(j\right)}\right]U_{1}\left(t\right)
\end{equation}
\[
\tilde{U}\left(t\right)=e^{-i\beta\bar{H}_{1}^{\left(j\right)}\left(t-jT_{0}\right)}\dots e^{-i\beta\bar{H}_{1}^{\left(0\right)}T_{0}}.
\]
If also the normal form transformation is performed the error is of
order $\beta^{\frac{3}{2}}$ .

\section{Iterative application and error analysis }

To reduce the error we introduce an iterative process. After the normal
form transformation is performed we introduce a new Hamiltonian
\begin{equation}
\tilde{H}_{1}=\frac{1}{\sqrt{\beta}}\tilde{U}_{1}^{-1}\left[H_{1}-H_{1}^{\left(g\right)}\right]\tilde{U}_{1}
\end{equation}
where
\begin{equation}
\tilde{U}_{1}=U_{0}U_{1}\tilde{U}
\end{equation}
The wave function at the new level satisfies a Schrödinger equation
like (3.12) of \cite{fishman2014multiscale} with $H_{1}$ playing
the role of $\tilde{A}\left(t\right)$ there, and $\beta$ is replaced
by $\beta^{3/2}$. Now one starts from an equation like \eqref{schred_pertr}
with $\beta$ replaced by $\beta^{3/2}$ and $H$ by $\tilde{H}_{1}$.
At each step the effective value of $\beta$ is reduced $\left(\beta\rightarrow\beta^{3/2}\right)$,
and so is the error. The process is repeated until the bound on the
error is satisfactory, as will be shown below. 

Assume the process repeated $l$ times. The resulting approximation
for the wave function is 
\begin{equation}
\psi\left(t\right)=\tilde{U}_{1}..........\tilde{U}_{l-1}\tilde{U}_{l}\psi\left(0\right)\label{eq:star}
\end{equation}
where $\tilde{U}_{l'}$ is the propagator at the $l'$ level of the
hierarchy, corresponding to the Hamiltonian $\tilde{H}_{l}$. 

We turn now to estimate rigorously the errors using the multi-scale
and averaging method, assuming $l$-levels of the hierarchy.\\
With $\beta$ small on a time interval of order $\beta^{\frac{1}{2}}\equiv T_{0}$.
Let us denote by $t_{max}$ the maximum time we want to simulate dynamics.
After introducing a normal form transformation we eliminate the $c\beta^{\frac{1}{2}}$
error term on (3.10) of \textbf{\cite{fishman2014multiscale}}, and
we get for the evolution 
\begin{equation}
\psi\left(t\right)=U\left(t\right)\psi\left(0\right)=U_{0}\left(t\right)R_{0}\left(t\right)\psi\left(0\right)
\end{equation}
leading to 
\begin{equation}
U(t)=U_{0}(t)R_{0}(t)\ \ \text{for}\ \ 0\leq t\leq t_{max}
\end{equation}
with $R_{0}(t)=1+\mathcal{O}\left(\beta^{\frac{3}{2}}t_{max}\right)$
for $0\leq t\leq t_{max}$ with $U_{0}(t)$ is given by (a product
of) averaged dynamics followed by a normal form unitary transformation.
Moreover 
\begin{equation}
i\frac{\partial R_{0}}{\partial t}=\beta^{\frac{3}{2}}H_{1}\left(t\right)R_{0}\left(t\right)\ ,\label{eq:R1eq}
\end{equation}
with 
\begin{equation}
\left\Vert H_{1}\left(t\right)\right\Vert =\mathcal{O}\left(1\right).
\end{equation}
If we then use the same method of averaging on $R_{0}(t)$ in Eq.
\eqref{eq:R1eq} we get $R_{0}(t)=U_{1}(t)R_{1}(t)$ and 
\begin{equation}
U\left(t\right)=\tilde{U}_{0}\left(t\right)U_{1}\left(t\right)R_{1}\left(t\right)\label{eq:58}
\end{equation}
where now $U_{1}(t)$ is the averaged approximate solution for $R_{0}(t),$
and 
\begin{equation}
R_{1}\left(t\right)=1+\mathcal{O}\left(\beta^{\frac{3}{2}}\right)+\mathcal{O}\left(\left(\beta^{\frac{3}{2}}\right)^{\frac{3}{2}}t_{max}\right).
\end{equation}
Again after normal form transformation, the $\mathcal{O}\left(\beta^{\frac{3}{2}}\right)$
correction drops and we have 
\begin{equation}
U\left(t\right)=\tilde{U}_{0}\left(t\right)\tilde{U}_{1}\left(t\right)+\mathcal{O}\left(\left(\beta^{\frac{3}{2}}\right)^{\frac{3}{2}}t_{max}\right).
\end{equation}
After $l$-such iterations, we get the exact solution 
\begin{equation}
U\left(t\right)=\tilde{U}_{0}\left(t\right)\tilde{U}_{1}\left(t\right)\cdots\tilde{U}_{l}\left(t\right)+\mathcal{O}\left(\beta^{\left(\frac{3}{2}\right)^{l}}t_{max}\right).
\end{equation}
So the convergence of the scheme is super exponentially fast, close
to the Newton type iteration.\\
We denote by $t_{max}$ the maximum time we want to simulate our dynamics.
The error in the MSA level $l$ of the hierarchy denoted by $\mathcal{\epsilon}$
can be written as 
\begin{equation}
\mathcal{\epsilon}=\beta^{\left(\frac{3}{2}\right)^{l}}t_{max}.\label{eq:error_epsilon}
\end{equation}
We can then invert this relation to obtain the number of desired hierarchy
steps $l$ for a given $\beta$ and desired error $\epsilon$,
\[
\]
\begin{equation}
l>\frac{\log\left(\frac{\log\epsilon-\log t_{\text{max}}}{\log\beta}\right)}{\log\left(\frac{3}{2}\right)}.
\end{equation}
in terms of $\tau_{max}=\beta t_{max}$ (see Eq. \ref{tau_scale})
\begin{equation}
l>\frac{\log\left(\frac{\log\epsilon-\log\tau_{\text{max}}}{\log\beta}+1\right)}{\log\left(\frac{3}{2}\right)}\label{eq:error_ther_bound}
\end{equation}
The error in the integral \eqref{eq:H_1_j} is given by \eqref{eq:del_I_t}
and the error in the diagonalization of the averaged Hamiltonian is
assumed to be small (see discussion in the end of Sec. 5).

\section{Numerical implementation}

The purpose of this section is to demonstrate that the multi-scale
averaging method developed in \cite{fishman2014multiscale} is superior
to the standard split step method as it is much faster and the bound
on the error can be estimated analytically. We will evolve the wave
function for the model presented in Sec.\textbf{ }\ref{sec:The-model-that}
with the approximate evolution operator of the multiscale and averaging
(MSA) method and compare the results to the ones found using the standard
split step method. In the zeroth order we evolve the wave function
with the help of (\ref{zero_order_prog}), with $U_{0}$ calculated
by (\ref{zero_order_prop}). \\
The diagonalization (\ref{diag_zero}) can also be performed once
and can be done in parallel for the various time intervals. In particular
the first order (\ref{first_order_prop}) requires to diagonlize $\bar{H}_{1}^{(j)}$
that in turn is given by the diagonalized $\bar{H}_{0}^{(j)}$ given
by \eqref{eq:H_0}. This enables to compute the $\alpha_{k_{1},k_{0}}$
of (\ref{alphas}). To obtain the first order MSA approximate dynamics,
we use the evolution operator as given by \eqref{eq:star}.

\begin{figure}[H]
\begin{centering}
\includegraphics[scale=0.35]{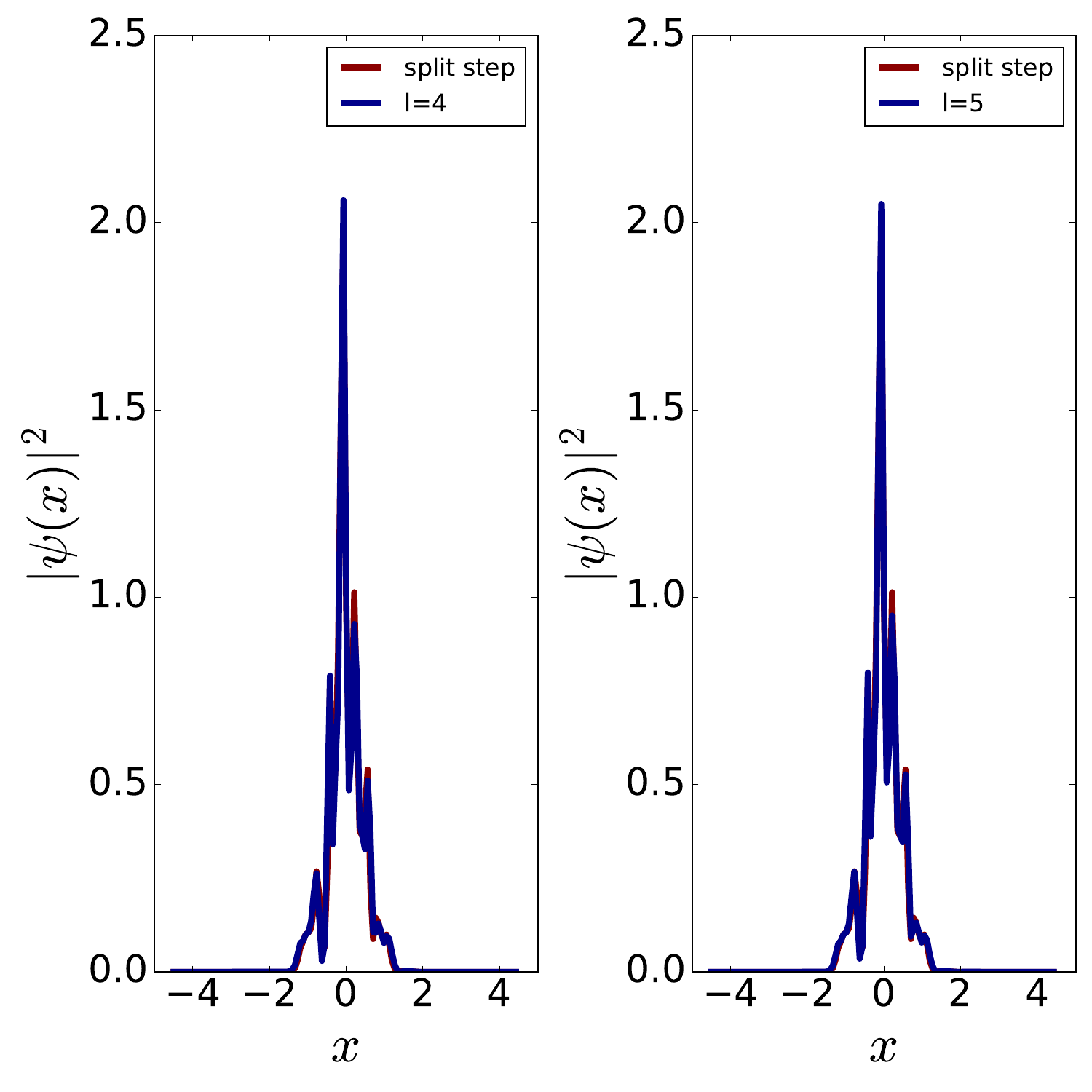}
\par\end{centering}

\protect\caption{\label{fig:The-wave-function}The wave function at time $t=T_{0}=\frac{1}{\sqrt{\beta}}$
obtained from the averaging method with precision of $\epsilon=10^{-7}$
and the split step method with accuracy of $\delta_{a}=10^{-7}$ for
$\beta=10^{-3}$ and $\sigma_{x}^{0}=1.0$. Two cases of multi-scale
and averaging are presented (a) The fourth level of the hierarchy
($l=4$, Eq. \ref{eq:error_epsilon}) (b) The fifth level of the hierarchy
($l=5$, Eq. \ref{eq:error_epsilon}). }
\end{figure}

\begin{figure}
\begin{centering}
\includegraphics[scale=0.3]{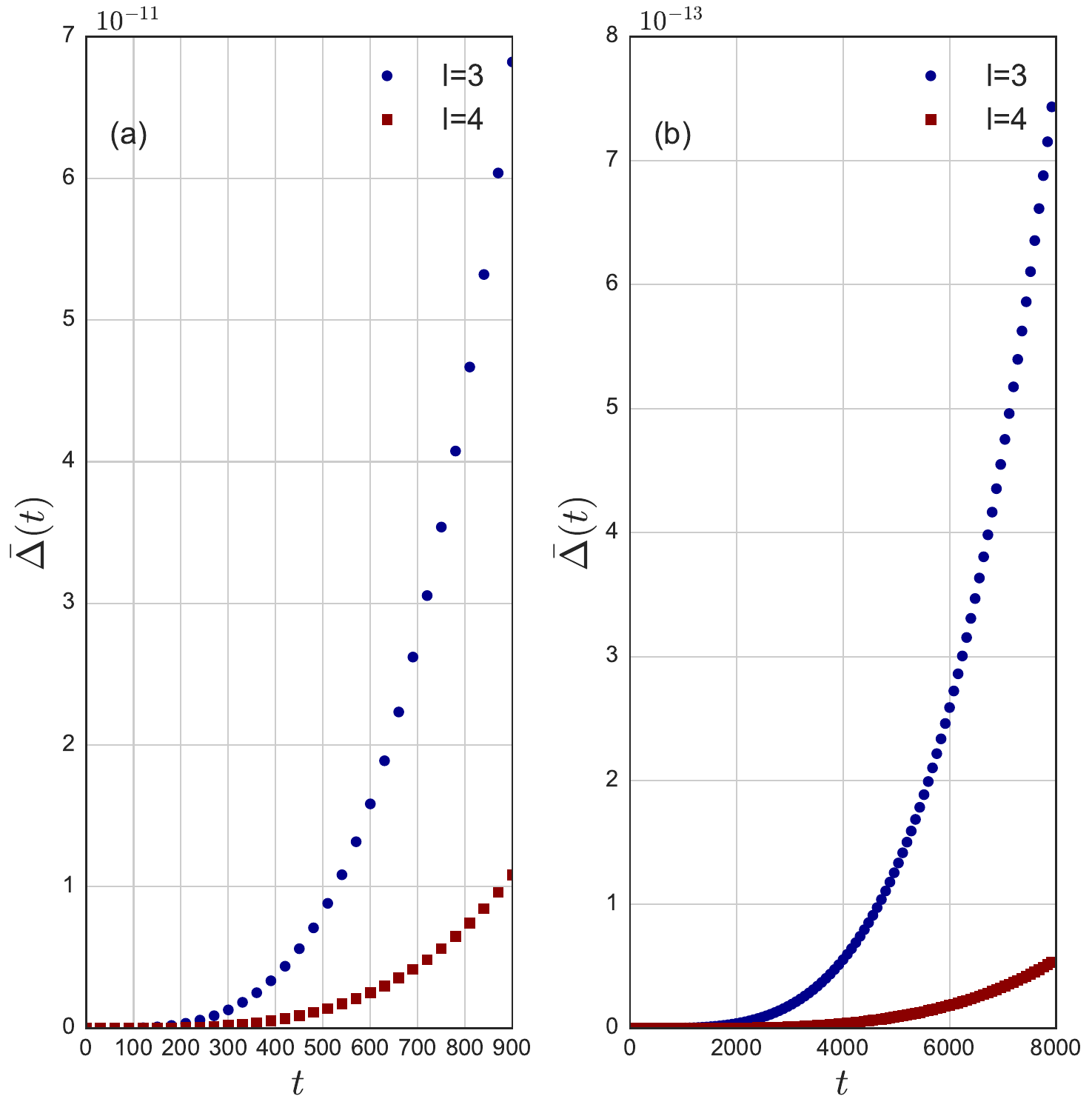}
\par\end{centering}

\protect\caption{\label{fig:The-difference1}The average difference $\bar{\Delta}\left(t\right)$
(Eqs. \ref{eq:delta_av}, \eqref{eq:error_diff}) between the split
step and averaging method as calculated for two different values of
(a) $\beta=10^{-3}$ and (b) $\beta=10^{-4}$ while $\sigma_{x}^{0}=1.0$.
Each subfigure presents both calculations for the third and fourth
level in the hierarchy ($l=3,4$ Eq. \ref{eq:error_epsilon}). The
precision required is $\epsilon=\delta_{a}=10^{-7}$ }
\end{figure}

The results are compared to the ones found with the help of the split
step method. In this method the wave function is propagated keeping
only the kinetic energy or the potential energy in small steps of
size $\delta t$. The value of the potential is taken in the center
of the time interval. The choice of a time step $\delta t$ is crucial.
The way to test the convergence of the scheme is by running the dynamics
up to a point $t$ using a time step $\delta t$, running the dynamics
up to the same point $t$ only using a new time step $\delta t'=\frac{1}{2}\delta t$.
If the wave function is the same within some fixed accuracy then the
scheme is assumed to be convergent. The accuracy is defined as 
\begin{equation}
\delta_{a}\left(t\right)=\int_{\Gamma}dx|\psi_{1}^{(\delta t)}(x,t)-\psi_{2}^{(\frac{\delta t}{2})}(x,t)|^{2}.\label{eq:delta_a}
\end{equation}
Where $\Gamma$ is the domain in space where the wave function is
defined. If the error $\delta_{a}\left(t\right)$ is not small then
one needs to continue adjusting until one converges. Listed in Table
\ref{split-step-time} are some values of the small parameter $\beta$
and the time scale it dictates. For longer time scale a smaller and
a more refined time step is needed in order to converge the split
step method. For the values listed in Table \ref{split-step-time}
an accuracy of $\delta_{a}=10^{-7}$ was chosen as a convergence criterion,
i.e if the two wave functions obtained at the same time $t$ with
different time steps $\delta t$ and $\frac{\delta t}{2}$ differed
by less then $\delta_{a}=10^{-7}$, the algorithm is considered converged
and an appropriate choice of $\delta t$ is obtained. The {*} marks
the fact that the required time was too long for the standard split-step
computation to converge.\\
To demonstrate the accuracy of the MSA method, we denote by $\psi_{a}(x,t)$
the wave function computed by the MSA method and by $\psi_{s}(x,t)$
the one found by the split step method and compute the deviation 
\begin{equation}
\Delta(t)=\int_{\Gamma}dx|\psi_{s}(x,t)-\psi_{a}(x,t)|^{2}\label{eq:error_diff}
\end{equation}
The initial wave function in all our computations is 
\begin{equation}
\psi\left(x,t=0\right)=\mathcal{N}e^{-\frac{x^{2}}{2\left(\sigma_{x}^{0}\right)^{2}}},\label{eq:init_wf}
\end{equation}
where $\mathcal{N}$ is the normalization constant. In Fig. \ref{fig:The-wave-function}
the comparison between the wave functions $\psi_{s}\left(x,t\right)$
and $\psi_{a}\left(x,t\right)$ is obtained by evolution starting
from the initial wave function \eqref{eq:init_wf}. It is presented
for an arbitrary realization of the random potential \eqref{eq:random potential in moving frame}.
The difference is very small and it will be calculated in what follows.
The results are presented in Fig \ref{fig:The-difference1} where
we plot 
\begin{equation}
\bar{\Delta}\left(t\right)=\left\langle \Delta\left(t\right)\right\rangle _{\text{av}}.\label{eq:delta_av}
\end{equation}
The average is over 40 realizations of the potential $V\left(x,t\right)$
of (11). The averaging is performed as follows: $\Delta\left(t\right)$
is calculated for a specific realization and the average is taken
so that the $\phi_{i}$ are distributed uniformly in the interval
$[-\pi,\pi]$, while $v_{n}^{(r)}$ and $k_{n}$ are distributed uniformly
in the intervals $[v^{min}=-15,v^{max}=15]$ and $[k^{min}=-20,k^{max}=20]$
respectively (see Sec. 2 paragraph following Eq. 5 ). $\omega_{n}$
is calculated by \eqref{eq:5_w}. We take $\beta=1\cdot10^{-3}$ and
$\beta=1\cdot10^{-4}$ while $\sigma_{x}^{0}=1$ resulting in the
initial momentum standard deviation $\sigma_{k}^{0}=0.5$. In our
basis Eqs. \eqref{diag_zero} and \eqref{eq:basis_wave} for\textbf{
$|\varphi_{j}^{k}>,$} the largest value of the momentum is $\left|\tilde{k}\right|=80$,
therefore 
\begin{equation}
\left|\tilde{k}\right|\gg\sigma_{k}^{0}.\label{eq:k_tilde}
\end{equation}
The largest value of $\left|x\right|$ is $10$ therefore $\sigma_{x}^{0}$
is much smaller then the largest value of $\left|x\right|$, namely
10. Note that $\Delta\left(t\right)$ is much smaller then $\delta_{a}$
and $\epsilon,$ indicating that the results are much more accurate
then expected from the theoretical bounds. This may be specific to
the potential we used. A similar situation was observed in the appendix
of \cite{fishman2014multiscale}. 

\begin{figure}
\begin{centering}
\includegraphics[scale=0.45]{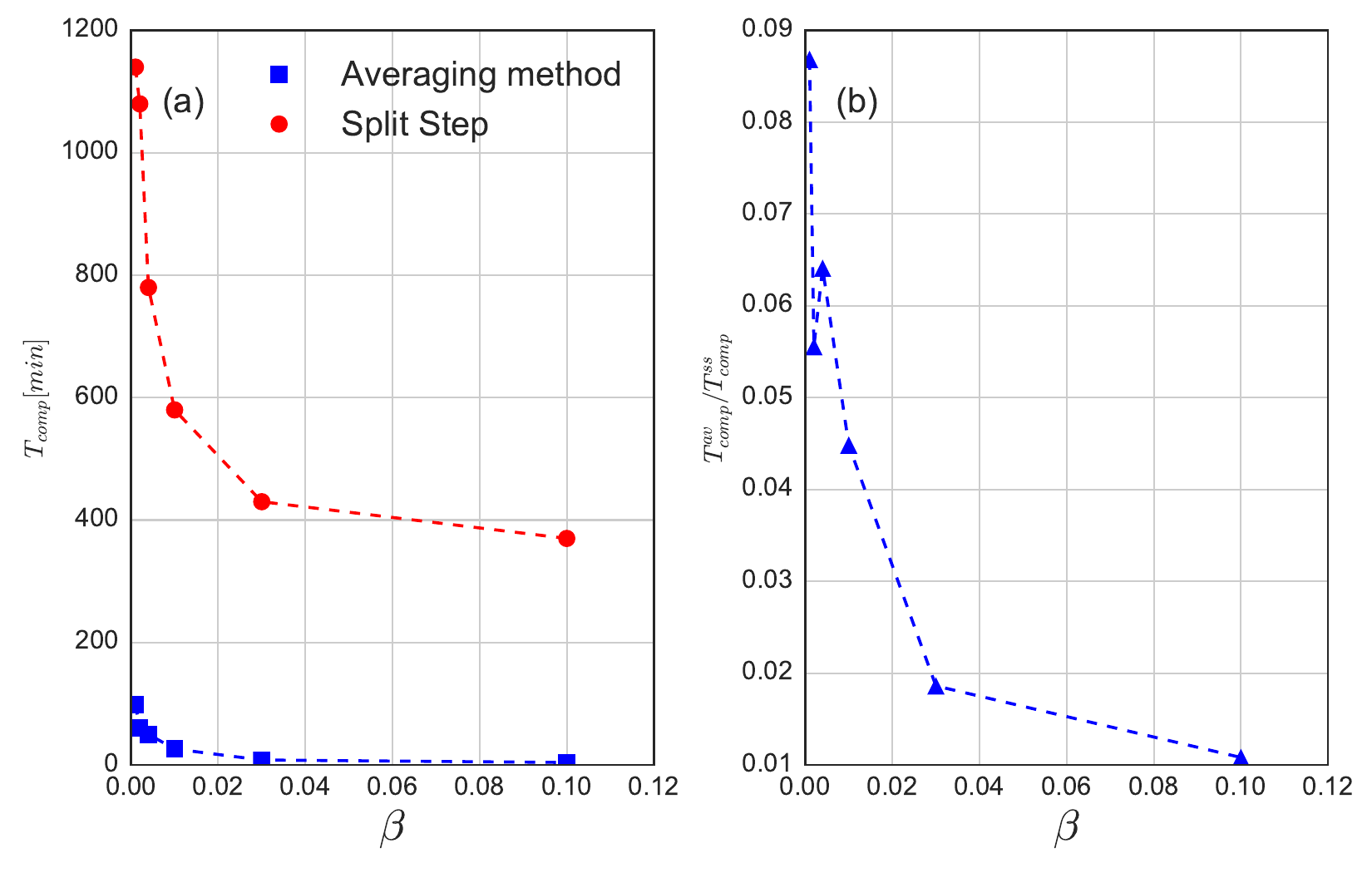}
\par\end{centering}

\protect\caption{\label{fig:Computational-ratio} Computational benchmark (run time)
as a function of $\beta$ as presented in Table \ref{tab:The-comparison-between}.
(a) Run times, obtained for split step method\textbf{ ($T_{comp}^{ss}$)}
and for the multiple scale and averaging ($T_{comp}^{av}$) method.
The precision required was $\epsilon=\delta_{a}=10^{-7}$ and the
values of $l$ are presented in table \ref{tab:The-comparison-between}.
(b) The ratio $\frac{T_{comp}^{av}}{T_{comp}^{ss}}$ presented in
table \ref{tab:The-comparison-between}.}
\end{figure}

To demonstrate the efficiency of the calculation we compare the computer
time $\text{T}_{\text{comp}}$ required to perform the numerical time
propagation Fig. \ref{tab:The-comparison-between} up to times $t_{max}=8000$,
we compare the results for the multi-scale and averaging split steps
methods with the same precision $\delta_{a}=\epsilon=10^{-7}$ we
choose a time $t_{max}^{'}=T_{0}\cdot\bar{i}$ such that $\bar{i}$
is the smallest integer satisfying $t_{max}^{'}>t_{max.}$ the lowest
hierarchy $l_{min}$ required for the calculation $l$ is used. The
results are summarized in table \ref{tab:The-comparison-between}
and plotted in Fig \ref{fig:Computational-ratio}. The calculations
were performed on two computational nodes each composed of a 2.4 Ghz
Intel Xeon processors.

\begin{table}[H]
\begin{centering}
\begin{tabular}{|c|c|c|c|c|c|}
\hline 
$\beta$  & $v$  & $T_{0}^{2}$  & $T_{0}^{4}$  & $\delta t$ for $T_{0}^{2}$  & $\delta t$ for $T_{0}^{4}$\tabularnewline
\hline 
\hline 
$1\cdot10^{-2}$  & $100$  & $100$  & $10000$  & $0.01$  & $0.0009$\tabularnewline
\hline 
$1\cdot10^{-3}$  & $1000$  & $1000$  & $1000000$  & $0.01$  & $0.0001$\tabularnewline
\hline 
$5\cdot10^{-4}$  & $2000$  & $2000$  & $400000$  & $0.009$  & $*$\tabularnewline
\hline 
$1\cdot10^{-4}$  & $10000$  & $10000$  & $100000000$  & $0.009$  & {*}\tabularnewline
\hline 
\end{tabular}
\par\end{centering}

\protect\protect\caption{\label{split-step-time}split step time steps for different values
of the small parameter. The accuracy is $\delta_{a}=10^{-7}$}
\end{table}

 It turns out that for the problem we studied, the results we obtained
are probably much better then the error estimate \eqref{eq:error_epsilon}.
To see this we present in table \ref{delta34} the difference 
\begin{equation}
\tilde{\Delta}_{l,l+1}\left(t\right)=\int_{\Gamma}dx|\psi_{l}(x,t_{max})-\psi_{l+1}(x,t_{max})|^{2}\label{eq:diff_l_l11}
\end{equation}
indicating the order of magnitude of the error as well as the bound
\eqref{eq:error_epsilon} for $l=3$ and $l=4$ for different values
of $\beta$ and $t_{max}=T_{0}^{2}=\frac{1}{\beta}$ Indeed this is
the case. The error in the calculation of the integrand \eqref{eq:H_1_j}
is given by \eqref{eq:del_I_t}. The reason is that the matrix consisting
of the $\alpha_{k_{i},k_{j}}$ is almost a unit matrix. For all of
our calculations we verified that 
\begin{equation}
\sum_{k_{j}\neq k_{i}}^{M=64}\left|\alpha_{k_{i},k_{j}}\right|<10^{-23}
\end{equation}
while
\[
\sum_{k_{j}=k_{i}}^{M=64}\left|\alpha_{k_{i},k_{j}}-1\right|<10^{-20}.
\]
The error in the diagonalization of the averaged Hamiltonian $\bar{H}_{j}^{\left(g\right)}$
is of the order of $10^{-50}$. The diagonalization is performed by
the lancos algorithm \cite{lehoucq1998arpack}.

\begin{table}[H]
\begin{centering}
\begin{tabular}{|c|c|c|c|c|c|c|}
\hline 
$\beta$ & $5\cdot10^{-5}$ & $1\cdot10^{-4}$ & $3\cdot10^{-3}$ & $1\cdot10^{-2}$ & $3\cdot10^{-2}$ & $1\cdot10^{-1}$\tabularnewline
\hline 
\hline 
$T_{0}$ & 141.42 & 100 & 18.25 & 10 & 5.77 & 3.162\tabularnewline
\hline 
$\begin{array}{c}
\\
\bar{i}
\end{array}$ & 57 & 81 & 439 & 801 & 1386 & 2530\tabularnewline
\hline 
$t_{max}^{'}$ & 8061.02 & 8100 & 8015 & 8010 & 8002.07 & 8000.56\tabularnewline
\hline 
$l_{min}$ & 3 & 3 & 4 & 5 & 5 & 6\tabularnewline
\hline 
$T_{comp}^{av}$ & 93 & 63 & 50 & 25 & 8 & 4\tabularnewline
\hline 
$T_{comp}^{ss}$ & 1140 & 1080 & 780 & 580 & 420 & 380\tabularnewline
\hline 
$\frac{T_{comp}^{av}}{T_{comp}^{ss}}$ & 0.081 & 0.058 & 0.064 & 0.043 & 0.019 & 0.011\tabularnewline
\hline 
\end{tabular}
\par\end{centering}

\protect\caption{\label{tab:The-comparison-between}The comparison between the computational
time $T_{comp}^{av}$ (in minutes) using the multi-scale averaging
method and the computational time $T_{comp}^{ss}$ using the split
step method. Averaging over 40 realizations similar to the ones used
in Fig. 2 was performed.}
\end{table}

\begin{table}[H]
\begin{centering}
\begin{tabular}{|c|c|c|c|c|c|c|}
\hline 
$\beta$ & $5\cdot10^{-5}$ & $1\cdot10^{-4}$ & $3\cdot10^{-3}$ & $1\cdot10^{-2}$ & $3\cdot10^{-2}$ & $1\cdot10^{-1}$\tabularnewline
\hline 
\hline 
$\tilde{\Delta}_{3,4}$ & $6\cdot10^{-11}$ & $1.04\cdot10^{-10}$ & $1.44\cdot10^{-10}$ & $1.81\cdot10^{-10}$ & $2.17\cdot10^{-12}$ & $2.5\cdot10^{-10}$\tabularnewline
\hline 
$\epsilon_{l=3}$ & $75\cdot10^{-8}$ & $3.88\cdot10^{-7}$ & $2.02\cdot10^{-6}$ & $1.78\cdot10^{-5}$ & $2.42\cdot10^{-4}$ & $2.21\cdot10^{-3}$\tabularnewline
\hline 
$\epsilon_{l=4}$ & $6.5\cdot10^{-10}$ & $1.08\cdot10^{-9}$ & $1.81\cdot10^{-9}$ & $7.5\cdot10^{-8}$ & $6.5\cdot10^{-7}$ & $8.66\cdot10^{-5}$\tabularnewline
\hline 
\end{tabular}
\par\end{centering}

\protect\protect\caption{\label{delta34}The difference \eqref{eq:diff_l_l11} and the bound
\eqref{eq:error_epsilon} for various values of $\beta$ and $t_{max}=T_{0}^{2}=\frac{1}{\beta}.$ }
\end{table}

\section{Spreading in k-space }

\begin{figure}
\begin{centering}
\includegraphics[scale=0.2]{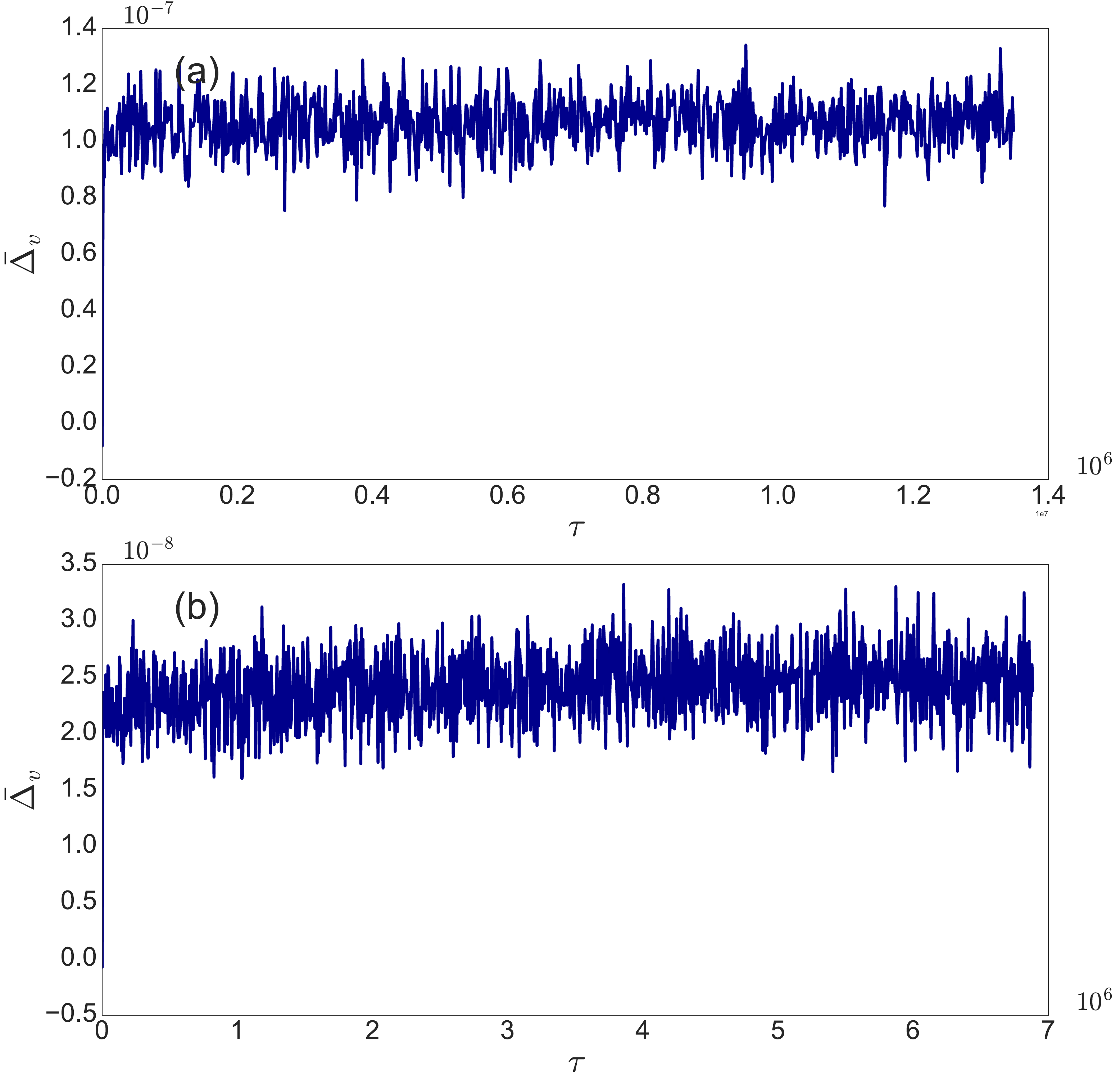}
\par\end{centering}

\protect\caption{\label{fig:Average-kinetic-energy}The average spread of the kinetic
energy $\bar{\Delta}_{v}$ of Eq. \eqref{eq:delta_bar_av} of a wave
function obtained for the multi-scale and averaging method with (a)
$\beta=5\cdot10^{-3}$ and $l=9$ (b) $\beta=1\cdot10^{-4}$ and $l=8$
\label{p23_f_4her}. For both cases $\epsilon=10^{-10}$. Note that
the hierarchy used here is much higher then required for the precision
$\epsilon$.}
\end{figure}

\begin{figure}
\begin{centering}
\includegraphics[scale=0.26]{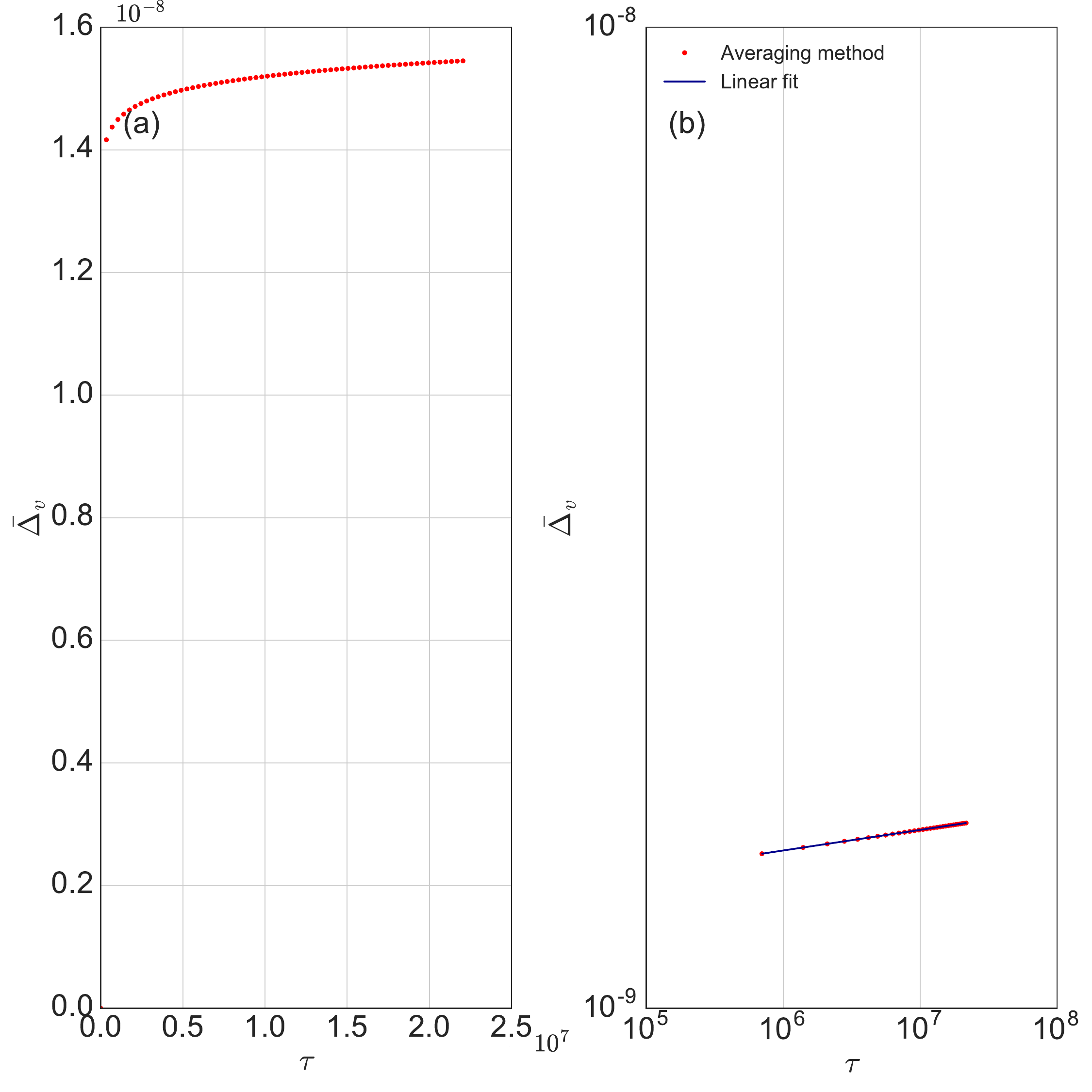}
\par\end{centering}

\protect\caption{$\bar{\Delta}_{v}\left(\tau\right)$ smoothened over time intervals
$\Delta\tau=10^{2}$ as a function of $\tau$ on (a) regular scale
(b) logarithmic scale, for $\beta=10^{-4}$, $\epsilon=10^{-10}$
and $l=6$ \label{spread_log_tau}.}
\end{figure}

\begin{figure}
\begin{centering}
\includegraphics[scale=0.25]{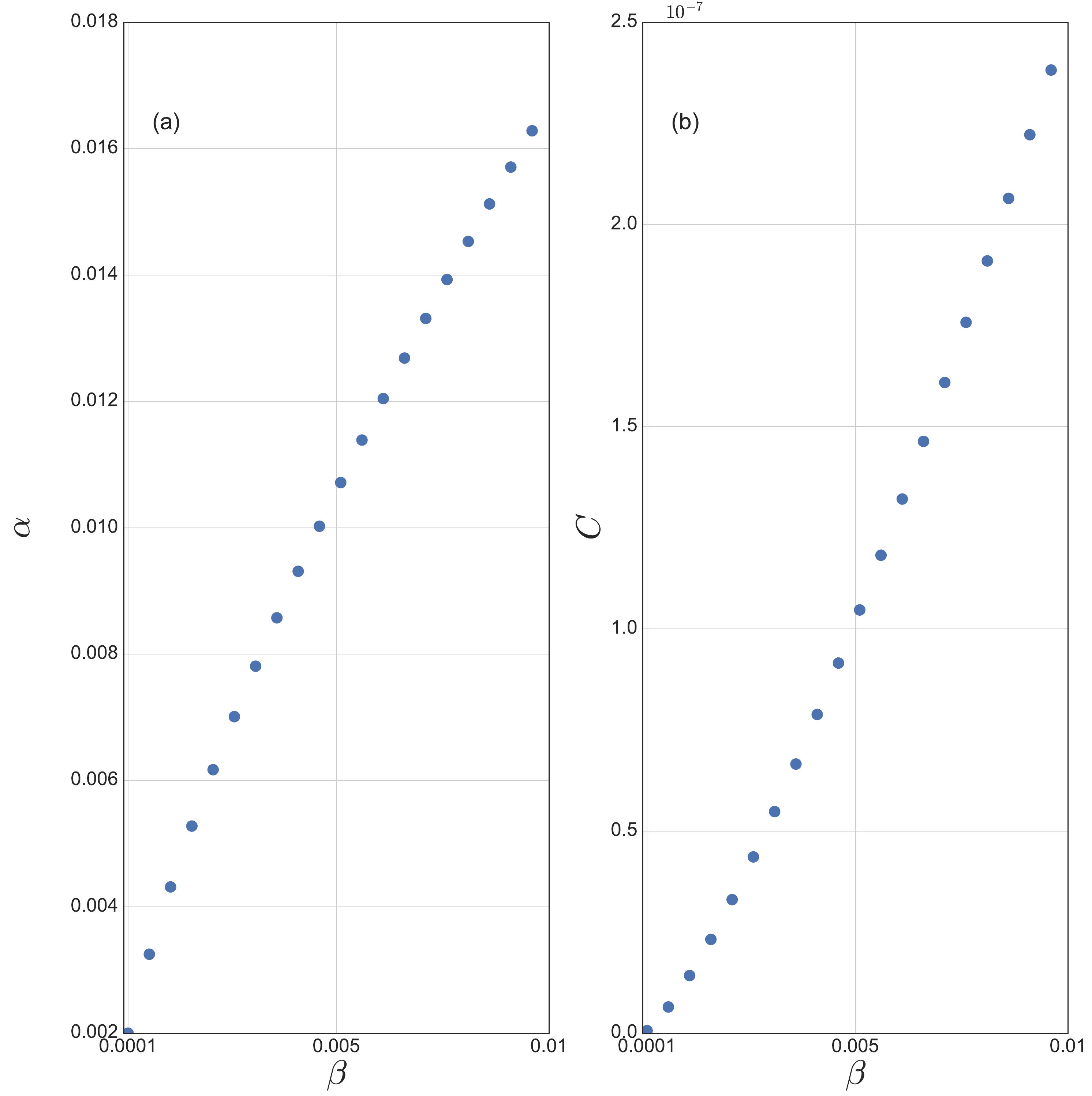}
\par\end{centering}

\protect\caption{The parameters of the fit \eqref{eq:fir_lin-1} as a function of $\beta$
\label{fit}}
\end{figure}

Multiscale and averaging (MSA) enables us to calculate very accurately
the spreading in momentum space over a very long time. For this purpose
we evolve the wave function $\psi\left(x,t\right)$ starting from
\eqref{eq:init_wf} with $\sigma_{x}^{0}=1$ for the potential \eqref{eq:random potential in moving frame}.
The wave function is used to calculate the variance of the momentum
\begin{equation}
\text{Var}_{k}\left(t\right)=\int\hat{\psi}^{*}\left(k,t\right)\left(k-\bar{k}\right)^{2}\hat{\psi}\left(k,t\right)dk\label{eq:var_k}
\end{equation}
 where 
\begin{equation}
\bar{k}=\int\hat{\psi}^{*}\left(k,t\right)k\hat{\psi}\left(k,t\right)dk
\end{equation}
 and $\hat{\psi}\left(k,t\right)$ is the Fourier transform of $\psi\left(x,t\right)$.
Then we calculate the spread of the momentum relative to initial one
\begin{equation}
\Delta\text{Var}_{k}^{\left(n\right)}\left(t\right)=\frac{\text{Var}_{k}\left(t\right)-\text{Var}_{k}\left(0\right)}{\text{Var}_{k}\left(0\right)}.
\end{equation}
This calculation is performed for each realization of the random potential.
Then average over 40 realizations of the random potential was performed
as in \eqref{eq:delta_av}, namely we calculate 
\begin{equation}
\bar{\Delta}_{v}\left(\tau\right)=\left\langle \Delta\text{Var}_{k}^{\left(n\right)}\left(t\right)\right\rangle _{\text{av}}\label{eq:delta_bar_av}
\end{equation}
 It is plotted as a function of $\tau=\beta t$ in Fig. \ref{fig:Average-kinetic-energy}.
Because of the smallness of $\bar{\Delta}_{v}$ we conclude that the
replacement of the Hamiltonian by a finite matrix does not effect
the result (see Eq. \eqref{eq:k_tilde}).  The plot is smoothed by
averaging over intervals of length $\Delta\tau=10^{2}$, leading to
the results presented in Fig. \ref{spread_log_tau}, the calculation
is repeated for several values of $\beta$, and in Fig. \ref{fit}
the results are fitted to the formula 
\begin{equation}
\bar{\Delta}_{v}\left(\tau\right)=C\left(\beta\right)\tau^{\alpha\left(\beta\right)}.\label{eq:fir_lin-1}
\end{equation}
From the plot it is reasonable to extrapolate 
\begin{equation}
\lim_{\beta\rightarrow0}C=\lim_{\beta\rightarrow0}\alpha=0.
\end{equation}
We conclude that in the limit $\beta\rightarrow0$ the spreading stops.
This is the limit where the velocity is much larger then the Chirikov
resonant velocities $v_{n}^{\left(r\right)}$ of Eq. \eqref{eq:5_w}.
It leads us to the conjecture that if in one dimension the $v_{n}^{\left(r\right)}$
are bounded, the kinetic energy cannot grow to infinity, in-spite
of the driving.

\begin{figure}
\includegraphics[scale=0.25]{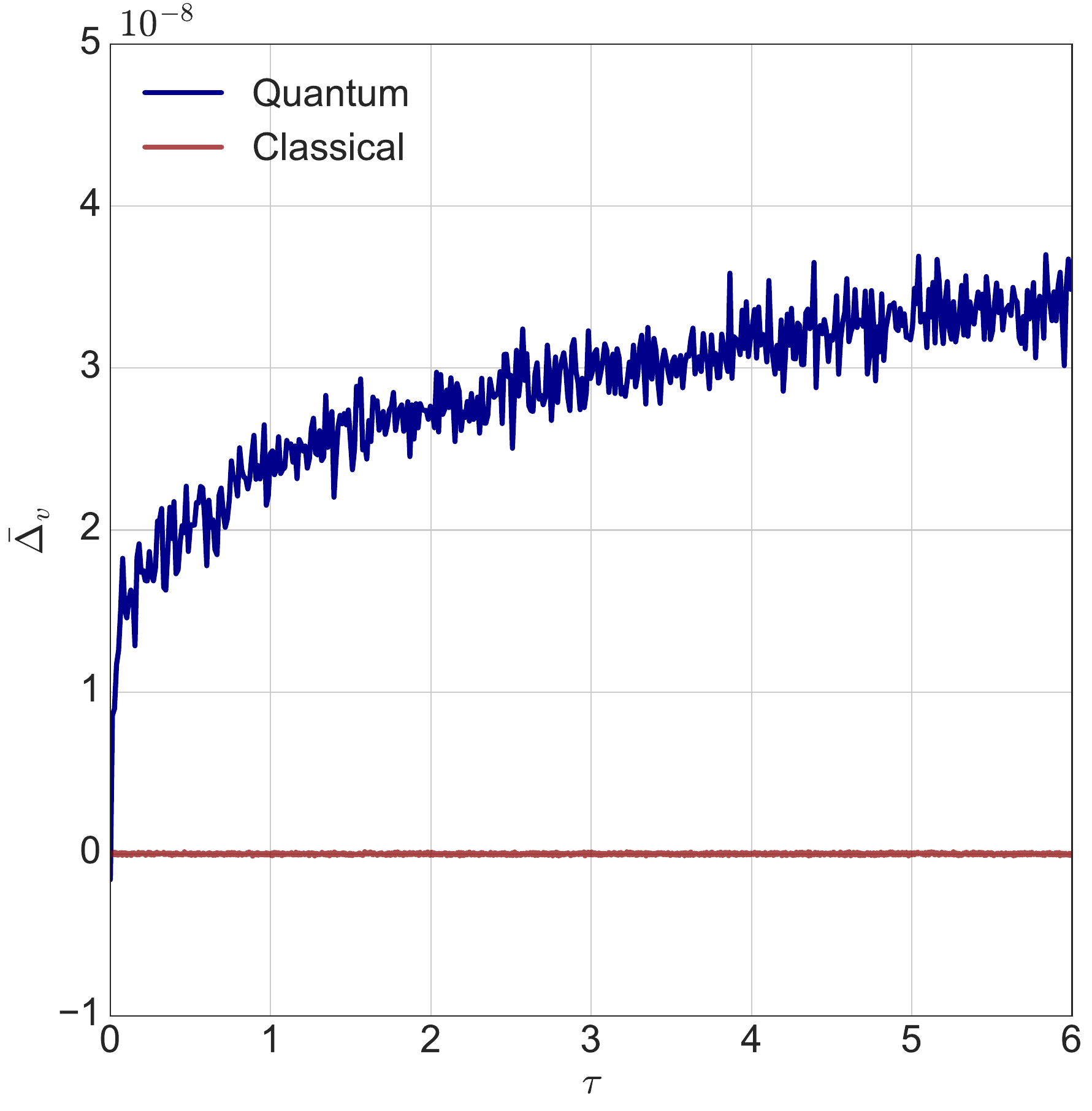}

\protect\caption{\label{fig:The-average-spreading}The average spreading of the quantum
and classical kinetic energy. The quantum results were found with
the multi-scale and averaging method and the classical result is obtained
via standard Runge Kutta integration with integration threshold of
$10^{-10}$. The parameters used are the same as in Fig. \ref{fig:Average-kinetic-energy},
but $l=6$ }
\end{figure}

In Fig. \ref{fig:The-average-spreading}, the classical and quantum
results are compared. The quantum results were computed as the ones
for Fig. \ref{fig:Average-kinetic-energy} while\textbf{ }for the
classical results 60 initial conditions were also chosen at random
from a Gaussian distribution corresponding to the initial quantum
wave function $\psi\left(x,0\right)$ of \eqref{eq:init_wf}. Both
classical and quantum results were averaged over 40 realizations of
the random potential. We note that both classical and quantum spreading
in momentum stops. The classical spreading stops at an earlier stage.

\section*{acknowledgments}

This work was partly supported by the Israel Science Foundation (ISF),
grant 1028/12, by the US-Israel Binational Science Foundation (BSF),
grant 2010132, by the USA National Science Foundation (NSF DMS 1201394)
and by the Shlomo Kaplansky academic chair. T.K. thanks the MIT physics
of living systems institute where part of this work was done. Part
of this work was done while T.K. and A.S. visited CCNU, China. T.K
acknowledges the grateful support of David Cohen and the ATLAS-Technion
grid project for computational resources.

\clearpage{}

\bibliographystyle{unsrt}
\bibliography{mybibfile2}

\end{document}